\documentstyle[twoside,fleqn,espcrc2]{article}

\newcommand{\eq}[1]{eq.(\ref{#1})}

\newcommand{\ben}{\begin{equation}}
\newcommand{\een}{\end{equation}}
\newcommand{\bea}{\begin{eqnarray}}
\newcommand{\eea}{\end{eqnarray}}
\newcommand{\bear}{\begin{array}}
\newcommand{\enar}{\end{array}}
\newcommand{\bdm}{\begin{displaymath}}
\newcommand{\edm}{\end{displaymath}}
\newcommand{\nn}{\nonumber \\ }
\newcommand{\binomial}[2]{\left (\begin{array}{c} {#1}\\ {#2} \end{array}
\right )}
\newcommand{\hf}{\frac{1}{2}}

\newcommand{\pa}{\partial}

\newcommand{\Z}{\mbox{$Z\hspace{-2mm}Z$}}

\newcommand{\br}{\langle}
\newcommand{\kt}{\rangle}
\newcommand{\bra}[1]{\langle {#1}|}
\newcommand{\ket}[1]{|{#1}\rangle}

\newcommand{\NP}[1]{Nucl.\ Phys.\ {\bf #1}}
\newcommand{\PL}[1]{Phys.\ Lett.\ {\bf #1}}
\newcommand{\CMP}[1]{Commun.\ Math.\ Phys.\ {\bf #1}}

\newcommand{\MPL}[1]{Mod.\ Phys.\ Lett.\ {\bf #1}}

\newcommand{\IM}[1]{Invent.\ Math.\ {\bf #1}}
\newcommand{\SJNP}[1]{Sov. J. Nucl. Phys.\ {\bf #1}}

\newcommand{\AmS}{{\protect\the\textfont2
  A\kern-.1667em\lower.5ex\hbox{M}\kern-.125emS}}

\title{
\begin{flushright}
NBI-HE-95-42\\
hep-th/9512175
\end{flushright}
Free field realization of $SL(2)$ correlators for admissible
representations, and hamiltonian reduction for correlators}

\author{J.L. Petersen\address{The Niels Bohr Institute, Blegdamsvej 17,
DK-2100 Copenhagen \O , Denmark}%
        \thanks{Talk presented by J.L. Petersen at the $29^{th}$ Symposium
Ahrenshop on the theory of elementary particles, Buckow, August 29 -- September
2, 1995},
        J. Rasmussen\address{The Niels Bohr Institute, University of
Copenhagen}
and M. Yu\address{The Niels Bohr Institute, University of Copenhagen,\\
address after March 1996: Inst. of Theor. Phys., Academia Sinica, Beijing,
Peoples Republic of China}}

\begin{document}

\begin{abstract}
A presentation is given of the free field realization relevant to $SL(2)$ WZW
theories with a Hilbert space based on admissible representations. It is
known that this implies the presence of two screening charges, one involving
a fractional power of a free field. We develop the use of fractional calculus
for treating in general such cases. We derive explicit integral representations
of $N$-point conformal blocks. We show that they satisfy the
Knizhnik-Zamolodchikov
equations and we prove how they are related to minimal conformal blocks via a
formulation of hamiltonian reduction advocated by Furlan, Ganchev, Paunov and
Petkova.
\end{abstract}

\maketitle

\section{INTRODUCTION}

In this talk I shall describe how to obtain conformal blocks for $SL(2)$
WZW theories in the case of non-integrable representations, in particular for
admissible representations \cite{KK,MFF}.
There are several reasons why this would be of
interest. First, a rather more complicated structure than is present for
integrable representations and in minimal models reveals itself, and it is
interesting to study in its own right. Second, our own principal interest
comes from refs.\cite{HY,AGSY} in which it has been demonstrated in
principle how
2-d quantum gravity coupled to minimal conformal matter may be described in
terms
of a topological $G/G$ model, with $G=SL(2)$. The way minimal conformal matter
arises is related to viewing these theories in terms of a hamiltonian reduction
of an $SL(2)$ theory \cite{Bel,Pol,BO}, and it is well known that this relation
depends on admissible representations. Finally, obtaining conformal blocks for
non-integrable representations may be interesting in various other ways, in
particular in relations to formulations of black hole string solutions as
discussed by Bars \cite{bars}.

There already exists a number of approaches towards the problem of conformal
blocks for admissible representations in the
literature \cite{BF,ATY,FGPP,D90,A,SV,FIM,FV,FF,MFF,An,AY,FM,FM95}.
However, our
goal here \cite{PRY} is to obtain a formulation based on the Wakimoto
\cite{Wak}
free field realization, and despite several attempts a complete solution has
so far been missing. The principal reason for this is related to the need
for introducing a second screening charge in the case of admissible
representations. This screening operator involves a fractional power of a free
(antighost) field \cite{BO}. Also the formalism requires the introduction of
several other fractional powers of free fields. It appears that this situation
cannot be resolved by bosonization. Instead I shall describe how everything may
be treated rather neatly by means of fractional calculus. The result is that we
appear to have a straightforward free field formalism.

As an explicit verification that our formalism works we have managed to prove
that the conformal blocks we obtain satisfy the Knizhnik-Zamolodchikov
equations
\cite{KZ,CF}. As an additional bonus we are able to provide a proof of an
interesting suggestion by Furlan, Ganchev, Paunov and Petkova \cite{FGPP} for
how conformal blocks of the $SL(2)$ WZW theory reduce to the conformal blocks
of minimal models. Even though the relation is often of a singular nature, the
result is very attractive.

When the level of the affine $\widehat{SL(2)}_k$ algebra is $k$, we define
\ben
k+2\equiv t=p/q
\een
where $p,q$ are positive coprime integers
for admissible representations. For these
there are degenerate representations whenever the spin is given by
\bea
2j_{r,s}+1&=&r-st\nn
r&=&1,...,p-1\nn
s&=&0,...,q-1
\eea
The relation to minimal conformal models is then as follows
\bea
h_{r,s+1}&=&\frac{1}{t}j_{r,s}(j_{r,s}+1)-j_{r,s}\nn
&=&\hf\alpha_{r,s+1}
(\alpha_{r,s+1}-2\alpha_0)\nn
\alpha_{r,s+1}&=&-j_{r,s}\sqrt{\frac{2}{t}}=\hf{[}(1-r)\alpha_+-s\alpha_-{]}\nn
\alpha_+&=&\sqrt{\frac{2}{t}}=-\frac{2}{\alpha_-}\nn
2\alpha_0&=&\alpha_++\alpha_-\nn
c&=&1-12\alpha_0^2=1-\frac{6(p-q)^2}{pq}
\label{minimal}
\eea
where $h$ is the conformal dimension and $c$ is the central charge for the
minimal model. So we see the need for understanding the case of non-integer
levels and spins.

\section{THE FREE FIELD REALIZATION}
The Wakimoto free field realization is in terms of a scalar field, $\varphi$
and a pair of bosonic dimension $(1,0)$ ghosts, $(\beta,\gamma)$:
\bea
\varphi(z)\varphi(w)&\sim&\log(z-w)\nn
\beta(z)\gamma(w)&\sim&\frac{1}{z-w}\nn
J^+(z)&=&\beta(z)\nn
J^3(z)&=&-\gamma\beta(z)-\sqrt{\frac{t}{2}}\pa\varphi(z)\nn
J^-(z)&=&-\gamma^2\beta(z)+k\pa\gamma(z)\nn
&-&\sqrt{2t}\gamma\pa\varphi(z)
\eea
They satisfy
\bea
J^+(z)J^-(w)&\sim&\frac{2}{z-w}J^3(w)+\frac{k}{(z-w)^2}\nn
J^3(z)J^\pm(w)&\sim&\pm\frac{1}{z-w}J^\pm(w)\nn
J^3(z)J^3(w)&\sim&\frac{k/2}{(z-w)^2}\nn
c&=&\frac{3k}{t}
\eea
Fateev and Zamolodchikov \cite{FZ} introduced a very useful formalism for
primary fields, which we shall adopt. In general there is a multiplet of
primary
fields $\phi_j^m(z)$. We combine these introducing an extra variable, $x$ as
follows
\ben
\phi_j(z,x)=\sum_m\phi_j^m(z)x^{j-m}
\een
For integrable representations $2j$ is an integer and we simply get a
polynomial
in $x$. However, for fractional spins we have highest weight, lowest weight or
continuous representations, and the $x$-dependence can be arbitrarily
complicated. The new primary field satisfies the following OPE
\bea
J^a(z)\phi_j(w,x)&\sim&\frac{1}{z-w}D_x^a\phi_j(w,x)\nn
D_x^+&=&-x^2\pa_x+2xj\nn
D_x^3&=&-x\pa_x+j\nn
D_x^-&=&\pa_x
\eea
Correlators now have an additional projective invariance related to the $x$
variable, thus the 3-point function satisfies
\bea
&&\br\phi_{j_3}(z_3,x_3)\phi_{j_2}(z_2,x_2)\phi_{j_1}(z_1,x_1)\kt\nn
&=&C_{123}\frac{(x_2-x_1)^{j_1+j_2-j_3}(x_2-x_3)^{j_2+j_3-j_1}}
{(z_2-z_1)^{h_1+h_2-h_3}(z_2-z_3)^{h_2+h_3-h_1}
}\nn
&\times&\frac{(x_1-x_3)^{j_1+j_3-j_2}}{(z_1-z_3)^{h_1+h_3-h_2}}
\label{projective}
\eea
Notice that for $x_i=z_i$ this reduces to the 3-point function of the
corresponding minimal model operators by virtue of \eq{minimal}. The
observation of \cite{FGPP} was that a similar situation seems to occur for any
$N$-point conformal block. This we shall prove towards the end. One easily
verifies that
\ben
\phi_j(z,x)={[}1+x\gamma(z){]}^{2j}e^{-j\sqrt{\frac{2}{t}}\varphi(z)}
\label{primary}
\een
Finally there are the two screening charge currents \cite{BO}
\bea
S_1(z)&=&\beta(z)e^{\sqrt{\frac{2}{t}}\varphi(z)}\nn
S_{-t}(z)&=&\beta(z)^{-t}e^{-t\sqrt{\frac{2}{t}}\varphi(z)}
\label{screening}
\eea
We see in these last equations the need for being able to treat fractional
powers of free fields.

\subsection{Fractional calculus}
Our treatment of Wick contractions is based on the
following identity, trivially valid for $-t$ a positive integer, but
non-trivial
for general $t$:
\bea
&&\beta(z)^{-t}F(\gamma(w))\nn
&=&:{[}\beta(z)+\frac{1}{z-w}\pa_{\gamma(w)}{]}^{-t}
F(\gamma(w)):\nn
&=&\sum_{n\in\Z}\binomial{-t}{n}:\beta^{n}(z)(z-w)^{t+n}\nn
&&\pa^{-t-n}_{\gamma(w)}F(\gamma(w)):
\eea
Examples of the use of fractional calculus \cite{MR} are the Riemann-Liouville
operator
\ben
\pa^{-a}f(z)=\frac{1}{\Gamma(a)}\int_0^z(z-t)^{a-1}f(t)dt, \ a>0
\een
and
\ben
\pa_x^a x^b=\frac{\Gamma(b+1)}{\Gamma(b-a+1)}x^{b-a}
\een
In addition we shall need unconventional (asymptotic) expansions like
\ben
e^x=\pa_x^ae^x=\sum_{n\in\Z}\frac{1}{\Gamma(n-a+1)}x^{n-a}
\een
for $x$ an operator.

\section{CONFORMAL BLOCKS}
According to the above, we may treat the free field representation of an
$N$-point conformal block in terms of the following integral of a free field
correlator:
\bea
&&\bra{j_N}\prod_{n=2}^{N-1}{[}1+x_n\gamma(z_n){]}^{2j_n}
e^{-j_n\sqrt{\frac{2}{t}}\varphi(z_n)}\nn
&\times&\oint\prod_{k=1}^{s}\frac{dv_k}{2\pi i}\beta^{-t}(v_k)
e^{-t\sqrt{\frac{2}{t}}\varphi(v_k)}\nn
&\times&\oint\prod_{l=1}^{r}\frac{dw_l}{2\pi i}\beta(w_l)
e^{\sqrt{\frac{2}{t}}\varphi(w_l)}\ket{j_1}
\label{npoint}
\eea
Here $r$ and $s$ are the number of screening charges of the first and
second kind respectively. I refer to \cite{PRY} for a discussion of the precise
choice of the bra and ket (see also below).
The notation implies that the corresponding primary
fields have been placed at $(z_1,x_1)=(0,0)$ and $(z_N,x_N)=(\infty,\infty)$,
thereby partly fixing the global $SL(2)$ invariances related to $z$ and to $x$.
We must now figure out how to do the contractions.

\subsection{The three point function}
In this case we have $j_1+j_2-j_3=r-st$ with $r,s$ being the number of
screening
operators. Due to projective invariance we know that the $x$ dependence of the
three point function will be of the form (cf. \eq{projective})
\ben
\bra{j_3}\{\phi_{j_2}(z,x)\}^{j_3}_{j_1}\ket{j_1}\propto
x^{r-st}
\een
Here $\{\phi_{j_2}(z,x)\}^{j_3}_{j_1}$ is the intertwining field defined by
means of the screening charges \cite{F}.
This tells us that we should expand
\bea
&&{[}1+x\gamma(z){]}^{2j_2}\nn
&=&\sum_{n\in\Z}\binomial{2j_2}{n+r-st}{[}x\gamma(z){]}^{n+r-st}
\eea
We also see that thus we are going to find the same net power of $\gamma$'s
as we have og $\beta$'s. One trick now is to employ
\bea
&&(1+x\gamma)^{2j}\nn
&=&\Gamma(2j+1)\oint\frac{du}{2\pi i}\frac{1}{u}
(\frac{D}{u})^{-2j}\exp(\frac{1+x\gamma}{u})\nn
&&\beta^a(w)e^{\frac{1+x\gamma(z)}{u}}\nn
&=&:(\beta(w)+\frac{1}{w-z}\pa_{\gamma(z)})^a
\exp{[}\frac{1+x\gamma(z)}{u}{]}:\nn
&=&:(\beta(w)+\frac{x/u}{w-z})^aD^a\exp(\frac{1+x\gamma(z)}{u}):
\eea
Here $D$ represents differentiation wrt the argument of the exponential
function.
Now it is relatively straightforward to write down the integral representation
of the three point function. The result is
\bea
W_3&=&\frac{\Gamma(2j_2+1)}{\Gamma(2j_2-r+st+1)}\int\prod_{i=1}^r\frac{dw_i}
{2\pi i}\prod_{j=1}^s\frac{dv_j}{2\pi i}\nn
&&\prod_{i_1<i_2}(w_{i_1}-w_{i_2})^{2/t}\prod_{j_1<j_2}(v_{j_1}-v_{j_2})^{2t}\nn
&&\prod_{i,j}(w_i-v_j)^{-2}\nn
&&\prod_{i=1}^rw_i^{(1-r_1)/t+s_1}(1-w_i)^{(1-r_2)/t+s_2-1}\nn
&&\prod_{j=1}^sv_j^{r_1-1-s_1t}(1-v_j)^{r_2-1-(s_2-1)t}
\eea
This result \cite{PRY} is a Dotsenko-Fateev integral
\cite{DF}, which may be analysed \cite{F} to provide the following fusion
rule ($2j_i+1=r_i-s_it$)
\bea
1+|r_1-r_2|\leq&r_3&\leq p-1-|r_1+r_2-p|\nn
|s_1-s_2|\leq&s_3&\leq q-1-|s_1+s_2-q+1|\nn
&&
\eea
This agrees with results in refs.\cite{AY,FM} referred to as their rule I.
However these authors also provide a rule II:
\bea
1+|p-r_1-r_2|\leq&r_3&\leq p-1-|r_1-r_2|\nn
1+|q-s_1-s_2-1|\leq&s_3&\leq q-2-|s_1-s_2|\nn
&&
\eea
This rule appears not to follow from our three point function. However, very
recently we have realized that (i) this rule is required in our 4-point
functions and (ii) may formally be derived from our 3-point function by
continuing to a negative number of screening charges. We intend to come back
elsewhere with a more detailed discussion.

\subsection{The general $N$-point block on the sphere}
Using the techniques described it is possible to write down the general
$N$-point function with $M=r+s$ screening charges as follows
\bea
W_N&=&\int\prod_{i=1}^M\frac{dw_i}{2\pi i}W_N^\varphi W_N^{\beta\gamma}\nn
W_N^\varphi&=&\prod_{m<n}(z_m-z_n)^{2j_mj_n/t}\nn
&\times&\prod_{i=1}^M\prod_{m=1}^{N-1}(w_i-z_m)^{2k_ij_m/t}\nn
&\times&\prod_{i<j<M}(w_i-w_j)^{2k_ik_j/t}\nn
W^{\beta\gamma}&=&\int\prod_{m=2}^{N-1}\frac{du_m}{2\pi i}\Gamma(2j_m+1)\nn
&\cdot&u^{2j_m-1}e^{1/u_m}\prod_{i=1}^MB(w_i)^{-k_i}\nn
B(w_i)^{-k_i}&=&(\sum_{l=1}^{N-1}\frac{x_l/u_l}{w_i-z_l})^{-k_i}
\eea
where $z_1=x_1=0$.
Here the powers $k_i$ are $-1$ and $t$ respectively for screening charges of
the first and second kind. Notice that there are some simple rules for how to
construct the $\beta\gamma$ part of the contractions. In particular
\ben
\beta(w_i)^{-k_i}\rightarrow B(w_i)^{-k_i}
\een
The above integral representation for $N$-point blocks is our main result.

\section{CHECKS ON THE RESULT}
We have performed several consistency checks on the above result, some of which
I mention here. So far we have constructed the conformal block as a correlator
\ben
\bra{j_N}\phi_{N-1}(z_{N-1},x_{N-1})...\phi_{j_2}(z_2,x_2)\ket{j_1}
\een
Here the chiral vertex operators, $\phi_l(z_l,x_l)$ are to be understood as
screened intertwining fields. The precise choice of screening contours define
which particular conformal block we are considering. This construction
presupposes that the formalism is properly $SL(2)$ invariant both as far as
$z$ and $x$ variables go. However, we might also consider the construction
based on
\ben
\bra{0}\phi_{j_N}(z_N,x_N)...\phi_{j_1}(z_1,x_1)\ket{0}
\een
In the limits
\bea
z_N,x_N&\rightarrow&\infty\nn
z_1,x_1&\rightarrow&0
\eea
the two should agree. This we have checked. In fact the choice of dual states
and dual vacuum is not manifestly $SL(2)$ invariant \cite{PRY,FMS}.
We use bra and ket states with the following properties
\bea
\bra{0}0\kt&=&1\nn
\bra{0}\gamma_0&=&0\nn
\bra{0}\beta_0&\neq& 0\nn
\bra{j}&=&\bra{0}e^{j\sqrt{\frac{2}{t}}q_\varphi}
\eea
where $q_\varphi$ is the position operator for the scalar field, $\varphi$.
Hence we
should also check that it is possible to prove that, nevertheless, the
formalism
is $SL(2)$ invariant. This we have done.

As a particular example of associativity (and other formal properties) we have
checked in great detail that
\ben{[}\beta^a(z)\gamma^a(w){]}{[}\beta^b(z)\gamma^b(w){]}=\beta^{a+b}(z)
\gamma^{a+b}(w)
\een

Probably the most interesting and stringent test we have performed is to
present
a detailed proof that our correlators satisfy the Knizhnik-Zamolodchikov
equations \cite{KZ}.
This means that our formalism constitutes a very powerful technique
for generating solutions to these equations. The KZ equations may be written as
\ben
\{t\pa_{z_{m_0}}+2\sum_{m\neq m_0}\frac{D^a_{x_{m_0}}D^a_{x_m}}{z_m-z_{m_0}}\}
W_N=0
\een
One possible way of proving this equation is to consider the functions,
$G^a(w)$ and $G(w)$ defined for a correlator of operators ${\cal O}$ by
\bea
G^a(w)&=&\br J^a(w){\cal O}\kt\nn
G(w)&=&\frac{1}{w-z_{m_0}}\{D^+_{x_{m_0}}G^-(w)\nn
&+&2D^3_{x_{m_0}}G^3(w)+D^-_{x_{m_0}}G^+(w)\}
\eea
For
\ben
{\cal O}=\phi_{j_N}(z_N,x_N)...\phi_{j_1}(z_1,x_1)
\een
one easily sees that $G(w)$ only has pole singularities at points $w=z_m$ and
no
pole at infinity:
\ben
G(w)\sim{\cal O}(1/w^2)
\een
Then the condition
\ben
\sum_m\mbox{Res}G|_{w=z_m}=0
\een
is the KZ equation. In our case we may build the function $G(w)$ as well. If
our
formalism based on fractional calculus is guaranteed to have all the correct
associativity properties of the operator algebra, it is trivial that we should
find the
same result. However, that is what we want to check. Hence we build the
function $G(w)$ using our rules. We observe that it has pole singularities at
$w=z_m$, but in addition there are singularities also at $w=w_i$,
the positions
of the screening charge currents before they are integrated over. However, we
may prove that these residues are total derivatives so that those contributions
vanish. Finally the residues of $G(w)$ at $w=z_m$ turn out to be exactly the
different terms in the KZ equations for our conformal block. This completes
(a sketch of) the proof.

\section{HAMILTONIAN REDUCTION}
The $N$-point conformal block,
\ben
W(z_N,x_N,...,z_1,x_1)
\een
is a function of $N$
pairs of variables, $(z_i,x_i)$. The proposal of Furlan, Ganchev, Paunov and
Petkova \cite{FGPP} which we have alluded to in the introduction, is that when
we put
\ben
x_i=z_i
\een
then this block agrees up to normalisation with a corresponding block in the
minimal conformal theory with the same $p,q$ as for the $\widehat{SL(2)}_k$
theory with $t=k+2=p/q$. This statement was verified in many examples in ref.
\cite{FGPP}. Using our technique we are in a position to present a proof and
also to clarify how the result is related to the more standard version of
hamiltonian reduction, based on
\ben
J^+(z)\sim 1
\een
In addition we partly find a somewhat stronger result, partly we also find that
the factor of proportionality may easily become zero.
Further, the behaviour of the conformal block as
$x_i\rightarrow z_i$ can be non-holomorphic.
In these cases the program of
\cite{FGPP} would seem to be in difficulty. More precisely, we have found that
one has the following
\bea
&&W_N^{WZW}(\{z_i,x_i=x\cdot z_i\})\nn
&=&c_N(\{j_l\},x)W^{\mbox{minimal}}_N(\{z_i\})
\eea
and we have obtained the coefficient, $c_N$. The treatment is based on the
observation that since
\ben
{[}J^a_n,\phi_j(z,x){]}=z^nD_x^a\phi_j(z,x)
\een
then
\bea
\phi_j(z,x\cdot z)&=&e^{zxJ^-_0}\phi_j(z,0)e^{-zxJ^-_0}\nn
&=&e^{xzD^-_y}\phi_j(z,y)|_{y=0}\nn
&=&e^{xJ^-_1}\phi_j(z,0)e^{-xJ^-_1}\nn
&=&e^{xJ^-_1}:e^{-j\sqrt{\frac{2}{t}}\varphi(z)}:e^{-xJ^-_1}
\eea
We may then express the $N$-point conformal block as follows
\bea
W_N&=&\bra{j_N}\phi_{j_{N-1}}(z_{N-1},x\cdot z_{N-1})...\nn
&...&\phi_{j_2}(z_2,x\cdot z_2)
\prod_i\oint\frac{dw_i}{2\pi i}S_{k_i}(w_i)\ket{j_1}\nn
&=&\bra{j_N}e^{xJ^-_1}\phi_{j_{N-1}}(z_{N-1},0)...\phi_{j_2}(z_2,0)\nn
&&\prod_i\oint\frac{dw_i}{2\pi i}\beta(w_i)^{-k_i}
e^{-k_i\sqrt{\frac{2}{t}}\varphi(w_i)}\ket{j_1}
\eea
Formally this is completely straightforward, however there is a tricky mode
question which must be examined when identities like
\ben
e^{-xJ^-_1}e^{xJ^-_1}=1
\een
are used, since it turns out that the two exponentials in general require
different fractional expansions. However, all is well \cite{PRY}. The crucial
observation now is that it may be shown that
\bea
\bra{j_N}e^{xJ^-_1}&=&\bra{j_N}(1-x\gamma_1)^{k-2j_N}(-)^{-r+st}\nn
&\cdot&\frac{\sin\{\pi(k-2j_N)\}}
{\sin\{\pi(k-2j_N-r+st)\}}\nn
\eea
where again $r,s$ denote the number of screening operators of the first and
second kind respectively. We therefore see that the only $\gamma$ dependence
in the correlator is via the mode $\gamma_1$. This mode only interacts with the
$\beta_{-1}$ mode which in turn multiplies $w_i^0=1$. Thus the $\gamma\beta$
part in the above correlator is trivial and all the remaining ingredients are
exactly a free field representation of a minimal model correlator. This
completes our proof of the FGPP proposal \cite{FGPP}. Thus the formulation of
hamiltonian reduction at the level of conformal blocks has an extremely simple
realization: one has to put the $x$-variables in the $SL(2)$ correlator
proportional to the $z$ variables. In the next subsection we examine how
the result is related to more standard formulations of hamiltonian reduction.

\subsection{Relation to standard hamiltonian reduction}
In our case the standard formulation of hamiltonian reduction is in terms of
imposing the constraint
\ben
J^+(z)=1
\een
It is convenient to view this constraint as a two step process: (i) first, all
modes except $J^+_{-1}$ are not operative, this makes $J^+(z)$ act like a
constant, and (ii) the constant is put equal to a definite value, like $1$.
In our free field formulation, $J^+(z)$ is represented by $\beta(z)$. We see
that in order for a correlator to respect the constraint, only the mode
$J^+_{-1}=\beta_{-1}$ must be active. This in turn requires that the only
$\gamma$ dependence that can be tolerated is via $\gamma_1$ and hence via
$J^-_1$. We see that this is precisely the crucial property we observed above.
Thus the most general (chiral) correlator respecting the constraint will have
to be of the form
\bea
W_{\mbox{constr}}&=&\bra{j_N}F(J^-_1)\phi_{j_{N-1}}(z_{N-1},0)...\nn
&...&\phi_{j_2}(z_2,0)\prod_i\oint\frac{dw_i}{2\pi i}\beta(w_i)^{-k_i}\nn
&\cdot&e^{-k_i\sqrt{\frac{2}{t}}\varphi(w_i)}\ket{j_1}\nn
&=&\bra{j_N}f(\gamma_1)\phi_{j_{N-1}}(z_{N-1},0)...\nn
&...&\phi_{j_2}(z_2,0)\prod_i\oint\frac{dw_i}{2\pi i}\beta(w_i)^{-k_i}\nn
&\cdot&e^{-k_i\sqrt{\frac{2}{t}}\varphi(w_i)}\ket{j_1}
\eea
We have shown that our $SL(2)$ correlator with $x_i=x\cdot z_i$ is exactly of
this form. In order for the value of the constant to be $1$ it is required that
\ben
f(\gamma_1)=e^{-\gamma_1}
\een
This condition is not satisfied by our correlator,
hence there is a non trivial constant of
proportionality between the reduced $SL(2)$ correlator and the minimal model
one.

\section{OUTLOOK}
It appears that we have solved a rather non-trivial problem, namely that of
writing down $SL(2)$ conformal blocks for admissible representations.
Equivalently our technique contains a powerful method for generating solutions
of the Knizhnik-Zamolodchikov equations. We emphasise that all this is achieved
with the physically convenient free field technique, although it was required
to apply fractional calculus to do that.

Several more steps have to be taken before we can achieve our original goal of
treating 2-d quantum gravity with this technique and in particular generalise
that. Here we mention some of these steps most of which represent work in
progress.

We must understand the normalization of the three point functions in order to
calculate 2-d gravity dressed ones using the $SL(2)/SL(2)$ cohomology. This
requires us to extend the program of \cite{DF} to the present case. On this
problem we are currently working. Next we should generalise the free field
formalism to higher groups and supergroups. Several steps in this direction
have
already appeared in the literature, but we need to extend that. Here we have
some partial results. It would be interesting to understand the meaning of
the constraint $x_i=z_i$ on higher genus Riemann surfaces and indeed for higher
groups ($W$-like strings). These and more questions need be examined before
we can truly use $G/G$ theories as a general tool to study generalised
non-critical string theory.


\begin{thebibliography}{9}
\bibitem{KK} V.G. Kac and D.A. Kazhdan, Adv. Math. {\bf 34} (1979) 79
\bibitem{MFF} F.G. Malikov, B.L. Feigin and D.B. Fuks,
Funkt. Anal. Prilozhen {\bf 20} (1986) 25
\bibitem{HY} H.-L. Hu and M. Yu, \PL{B 289} (1992) 302;\\
H.-L. Hu and M. Yu, \NP{B 391} (1993) 389
\bibitem{AGSY} O. Aharony, O. Ganor, J. Sonnenschein and S. Yankielowicz,
\NP{B 399} (1993) 527
\bibitem{Bel} A.A. Belavin, in {\em Proc. of the second Yukawa Symposium,
Nishinomiya, Japan, Springer Proceedings in Physics, Vol. 31 (1988) 132}
\bibitem{Pol} A.M. Polyakov, in {\em Physics and Mathematics of Strings,}
Eds. L. Brink, D. Friedan and A.M. Polyakov (World Scientific, 1990)
\bibitem{BO} M. Bershadsky and H. Ooguri, \CMP{126} (1989) 49
\bibitem{bars}I. Bars, hep-th/9503205 preprint
\bibitem{BF} D. Bernard and G. Felder, \CMP{127} (1990) 145
\bibitem{ATY} H. Awata, A. Tsuchiya and Y. Yamada, \NP{B 365} (1991) 680;\\
H. Awata, KEK-TH-310/KEK Preprint 91-189, preprint
\bibitem{FGPP} P. Furlan, A.Ch. Ganchev, R. Paunov and V.B. Petkova,
\PL{B 267} (1991) 63;\\
P. Furlan, A.Ch. Ganchev, R. Paunov and V.B. Petkova,
\NP{B 394} (1993) 665;\\
A.Ch. Ganchev and V.B. Petkova, \PL{B 293} (1992) 56
\bibitem{D90} Vl.S. Dotsenko, \NP{B 338} (1990) 747;\\
Vl.S. Dotsenko, \NP{B 358} (1991) 547
\bibitem{A} K. Amoto, J. Math. Soc. Japan {\bf 39} (1987) 191
\bibitem{SV} V.V. Schechtman and A.N. Varchenko, \IM{106} (1991) 139
\bibitem{FIM} B. Feigin and F. Malikov, Adv. Sov. Math. {\bf 17} (1993) 15,
hep-th/9306137;\\
K. Iohara and F. Malikov, Mod. Phys. Lett. {\bf A 8} (1993) 3613
\bibitem{FV} G. Felder and A. Varchenko, hep-th/9502165, preprint;\\
G. Felder and C. Wieczerkowski, hep-th/9411004, preprint
\bibitem{FF} B.L. Feigin and E. Frenkel,  Lett. Math. Phys. {\bf 19} (1990) 307
\bibitem{An}O. Andreev, HUB-IEB-94/9 hep-th/9407180, preprint;
O. Andreev, Landau-95-TMP-1/hep-th/9504082, preprint
\bibitem{AY} H. Awata and Y. Yamada, \MPL{A7} (1992) 1185
\bibitem{FM} B. Feigin and F. Malikov, Lett. Math. Phys {\bf 31} (1994) 315
\bibitem{FM95} B. Feigin and F. Malikov, preprint q-alg/9511011
\bibitem{PRY} J.L. Petersen, J. Rasmussen and M. Yu,
NBI-HE-95-16/hep-th/9504127, preprint, to be published in Nucl. Phys. B\\
J.L. Petersen, J. Rasmussen and M. Yu, NBI-HE-95-19/
hep-th/9506180, preprint, to be published in Nuc. Phys. B\\
J.L. Petersen, J. Rasmussen and M. Yu, NBI-HE-95-33/hep-th/9510059, preprint,
to be published in {\em Proc. on EU network meeting on Gauge Theories, Applied
Supersymmetry and Quantum Gravity, Leuven, Belgium, July 1995}
\bibitem{Wak} M. Wakimoto, \CMP{104} (1986) 60
\bibitem{KZ} V. Knizhnik and A. Zamolodchikov, \NP{B 247} (1984) 83
\bibitem{CF} P. Christe and R. Flume, \NP{B 282} (1987) 466
\bibitem{FZ} V.A. Fateev and A.B. Zamolodchikov, \SJNP{43} (1986) 657
\bibitem{MR} A.C. McBride and G.F. Roach (eds.) {\em Fractional Calculus}
(Pitman Advanced Publishing Program) (Boston 1985);\\
S.G. Samko, A.A. Kilbas and O.L. Marichec, {\em Fractional Integrals and
Derivatives,} Gordon and Breach, Science Publishers (1993)
\bibitem{F} G. Felder, \NP{B 317} (1989) 215 {[}Erratum:
{\bf B 324} (1989) 548{]}
\bibitem{DF} Vl.S. Dotsenko and V.A. Fateev, \NP{B 240}{[}FS12{]} (1984) 312;\\
\NP{B 251}{[}FS13{]} (1985) 691
\bibitem{FMS} D. Friedan, E. Martinec and S. Shenker, \NP{B 271} (1986) 93
\end{thebibliography}
\end{document}